\documentclass{article}

\usepackage{arxiv}

\usepackage[utf8]{inputenc} 
\usepackage[T1]{fontenc}    
\usepackage{hyperref}       
\usepackage{url}            
\usepackage{booktabs}       
\usepackage{amsfonts}       
\usepackage{nicefrac}       
\usepackage{microtype}      
\usepackage{lipsum}
\usepackage{graphicx, xfrac, subcaption, multirow}

\title{A Comparative Analysis of Relativistic Particle Pushers vis-à-vis Computation Time \& Accuracy}
\author{
 Mohammad Yasir \\
  Department of Physics,\\
  Indian Institute of Technology Delhi\\
  \texttt{yasir.iitd@outlook.com} \\
   \And
 Vikrant Saxena \\
  Department of Physics,\\
  Indian Institute of Technology Delhi\\
  \texttt{vsaxena@physics.iitd.ac.in} 
}

\begin{document}
\maketitle
\begin{abstract}
The performance of relativistic particle pushers has long been a topic of interest in the field of computational plasma physics, particularly from the point of view of the particle-in-cell approach. Previous works undertaken to compare such integrators have predominantly targeted the ultra-relativistic regime. In this paper, we utilize a custom-built code to study the core run-times of the Boris, the Vay, and the Higuera-Cary particle pushers for low-, high-, and ultra-relativistic particles. This is followed by a comparison of the three integrators in terms of accuracy and error. A fitness parameter is then proposed that can serve as a one-stop value to determine which method is more suitable for a particular simulation setup. It is hoped that through knowledge of such intricacies, the choice for the integrator will be easier to make depending on the problem at hand.
\end{abstract}

\section{Introduction}
\label{intro}
Relativistic plasma dynamics is a particularly engaging and versatile field, especially owing to the intricate and detailed calculations required for a comprehensive study of various such systems. Despite the rapid advancements in computing technologies, the possibility of first-principle calculations of the complete evolution of plasma systems still evades the scientific community. This limitation is owed primarily to the bottlenecks that remain in the form of computational resources available as regards to time and memory. Further, plasma systems ship with the additional complexity of multiple time, energy, and length scales that typically each require dedicated ways of study. Their spatial scales span a range of $10^8$, while the time scales of plasma processes can be anywhere between a few femtoseconds to a few days \cite{Tang_2005}. Even beyond that range, attosecond laser pulse generation is possible using plasma mirrors \cite{Thaury_2010}. Thus, there is no one-size-fits-all solution one can employ that is tractable, possible, and simultaneously, sufficiently accurate.

It should be noted that trajectory calculations, while discussed solely in the context of plasma simulations in this paper, aren't limited to that realm. Indeed, they form the backbone of a large class of scientific study. Common examples include molecular dynamics simulations \cite{allen2004introduction} and the study of the trajectories of extra-terrestrial objects (which, incidentally, is also often limited in terms of generality of solutions. For instance, the general three-body problem is still unsolved and recent developments have targeted restricted or planar restricted case only \cite{joung20243bp,guardia2024pcrtbp}). Thus, the accuracy, efficiency, and optimization of trajectory calculators is a topic of great study and is significant in range, number, as well as applications.

In the world of plasma simulations as well, the particle push often emerges as an expensive step. Fortunately, leapfrog integrators provide explicit, comparatively inexpensive, second-order ways of performing this step. Of these, perhaps the most well-known is the algorithm owed to J. P. Boris \cite{Boris1971ProceedingsFC}. As a near-symplectic \cite{zafarboris} integrator with excellent accuracy, it has served as the \textit{de-facto} standard in several widely accepted simulation software, e.g., EPOCH \cite{epoch}, SMILEI \cite{smilei}, etc. Other famously used algorithms include the Vay integrator \cite{Vay2007SimulationOB} and the Higuera-Cary pusher \cite{Hig2017Integrator}. There are also implicit integrators that involve more expensive computation but are less prone to errors at larger step sizes and thus, the computation time can be balanced with the number of iterations needed for the process under study to become fully resolved. Cohen \textit{et al.} \cite{cohen1982implicit} have proposed one such algorithm, as have ref \cite{Ripperda2017ACC}. Another method the reader can refer to is owed to J. P{\'e}tri \cite{petri2017fully}. Finally, a recently developed method that aims to preserve volume, energy, and the Lorentz invariance property in strong magnetic fields is owed to Zhang \textit{et al.} \cite{zhang2024structure}

In this paper, we focus on three explicit integrators viz., Boris, Vay, \& Higuera-Cary, and compare their performance vis-à-vis computation time and accuracy. It might be worth mentioning that Ripperda \textit{et al.} \cite{Ripperda2017ACC} have previously undertaken this task for the ultra-relativistic regime. Ref.  \cite{Vay2007SimulationOB,Hig2017Integrator} also perform this comparison, for more targeted scenarios. Our aim is a comprehensive study of the performance of these pushers for all three regimes of particle velocities, i.e. low-, high-, and ultra-relativistic (subsequently abbreviated as LR, HR, and UR).  The energy conservation of each of these methods has been well-studied by refs \cite{Ripperda2017ACC,Hig2017Integrator}. It is well-known that energy conservation is always upheld for the three integrators when an electric field is not present \cite{Hig2017Integrator}. A deep dive into the energy behaviour of Boris push was also done by Ref \cite{hairer2018energy}. The cross-field drift is best captured only by HC and Vay integrators \cite{Hig2017Integrator}. Hence, we restate that the search for a one-size-fits-all solution seems, at least for the moment, ill-fated. The hope with the present text is that with knowledge of the performance of pushers, a computational plasma physicist can make the more suitable choice while setting up their simulation.

To that end, we have developed a custom charged-particle tracking code titled \textbf{PaTriC - Particle Tracker in C++}, and used it to perform single-particle simulations with each of these methods. Opting for this model yielded us the ease of faster computations and rigorous analysis of the trajectory. We report our data in SI units, and start with a mathematical background in section \ref{background}. Section \ref{times} discusses core time of the integrators. Sections \ref{low_rel}, \ref{high_rel}, and \ref{ultra_rel} present the results we obtained in the LR, HR, and UR regimes. In section \ref{accuracy_param}, we develop a fitness parameter and finally, section \ref{conclusion} presents our closing remarks.

\section{Mathematical Background}
\label{background}

\subsection{Floating Point Operations and Vector Operations}
Of the many factors that affect the performance of an integrator in terms of run-time, the number of floating point operations (FLOPs) required at each iteration is perhaps the strongest contender. Addition and subtraction are typically inexpensive, well-optimized operations in all modern computational units and have negligible latency. Multiplication is slightly more expensive while division has traditionally been considered a slow operation \cite{oberman1997flop}. Luckily, much work has been undertaken to mitigate this discrepancy over time (Refs \cite{srinivas-1999,Schulte2007FloatingpointDA}). For the context of this paper, we will talk merely in terms of the absolute number of FLOPs required at each iteration. Furthermore, the square root operation, while considerably more expensive than simple addition and subtraction, will still be considered as a single FLOP for the ease of comparison.

A summary of the computational cost of vector operations in terms of floating point operations (FLOPs) is given in table \ref{tbl_vector_ops_cost}. A three-dimensional vector can be stored as an array of three floating point values. The efficiency of array storage and access has been a topic of great study\cite{logovzar2024exploring} but any cost incurred in implementation is far outweighed by the convenience that these structures offer. Also, the absolute number of FLOPs is not a true indication of the computational time. In practice, the performance will be affected by other factors like the exact nature of operations, the thermal state of the processor, the architecture of the processing unit, and the design of the program, among others. All the data we report are using our implementation which may differ slightly from others'.
\begin{table}
\centering

    \caption{\label{tbl_vector_ops_cost}Cost of different vector operations}
    \begin{tabular}{|p{0.3\textwidth}|c|c|}
        \hline
        \textbf{Operation}& \phantom{\quad} \textbf{Expression} \phantom{\quad}& \phantom{\quad} \textbf{Cost} (FLOPs)\phantom{\quad} \\
        \hline
        Addition& $\vec{a} + \vec{b}$& $3$\\
        Subtraction& $\vec{a} - \vec{b}$& $3$\\
        Scalar Multiplication& $\vec{a} \cdot \vec{b}$& $5$\\
        Vector Multiplication& $\vec{a} \times \vec{b}$& $9$\\
        Multiplication by scalar& $f\ \vec{a}$& $3$\\
        Division by scalar& $\sfrac{\vec{a}}{f}$& $3$\\
        \hline
    \end{tabular}
\end{table}

\subsection{Leapfrog Integrators}
\begin{figure}
\centering
\includegraphics[width=0.7\linewidth]{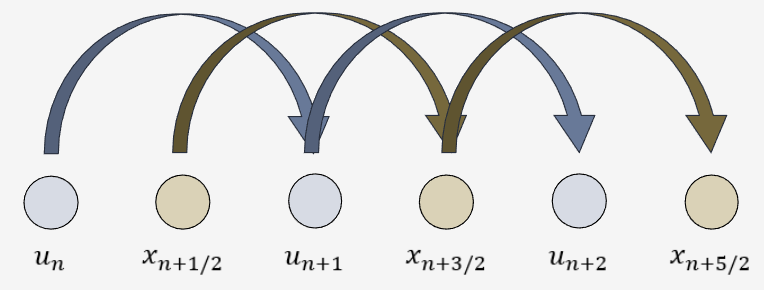}
\caption{\label{leapfrog}Demonstrative graphics for the staggered updates followed by leapfrog integrators.}%
\end{figure}

When performing the particle push, two essential properties of any integrator are time-reversibility and symplectic nature. It is possible to come up with a variety of algorithms that conform to these criteria. However, focus on simplicity yields the ones known as leapfrog integrators \cite{hockneyeastwood}, which stagger the process of position and velocity calculation (see figure \ref{leapfrog}). In this section, we present a summary of how leapfrog integrators work viz. the Boris, Vay, and the Higuera-Cary (abbreviated as HC for the remainder of this text) pushers.

Since we are dealing with the context of relativistic charged particle tracking, the equation of motion reads, with $\vec{u} = \gamma \vec{v}$, $\gamma = 1/\sqrt{1 - (v/c)^2}$ 
\begin{eqnarray}
    \label{vel_update}
    \frac{d\vec{u}}{dt} &=& \frac{q}{m} (\vec{E} + \vec{\bar{v}} \times \vec{B})\\
    \frac{d\vec{x}}{dt} &=& \vec{\bar{v}}
    \label{pos_update}
\end{eqnarray}
where $\vec{\bar{v}}$ represents some form of an average of the velocity between two time instances. When discretized, equations \ref{vel_update} and \ref{pos_update} looks as follows:
\begin{eqnarray}
\label{eqn_lorentz_discretized}
    \vec{u}_{i+1} &=& \vec{u}_i + (q/m) (\vec{E}_{i+\sfrac{1}{2}} + \vec{\bar{v}} \times \vec{B}_{i+\sfrac{1}{2}})\ dt\\
    \vec{x}_{i+\sfrac{1}{2}} &=& \vec{x}_{i-\sfrac{1}{2}} + \vec{\bar{v}}_i\ dt\ .
\end{eqnarray}

It is the choice of the form of $\vec{\bar{v}}$ in equation \ref{eqn_lorentz_discretized} that yields us different algorithms for trajectory calculation, each with their own advantages and disadvantages. Usually, the aim is to hit the perfect balance of accuracy, generality, and speed. In section \ref{accuracy_param}, we will attempt to propose a parameter that can summarise all the factors.

\textbf{Quick Note}: Through the rest of this text, we utilize the following common notation.
\begin{eqnarray}
    f &=& \frac{q\delta t}{2m}\\
    \vec{\epsilon} &=& f\vec{E}\\
    \vec{\beta} &=& f\vec{B}\\
    \Gamma(\vec{u}) &=& \sqrt{1 + \frac{\vec{u} \cdot \vec{u}}{c^2}} \text{, $\vec{u}$ is any vector.}
\end{eqnarray}
Furthermore, our program has been designed for constant time-stepping, With adaptive time-stepping, the results are bound to change since it has been demonstrated \cite{skeel1992symplectic} that variable time-stepping will destroy the symplectic nature of integrators been designed for a fixed step-size. Hence, adaptive time-stepping algorithms and variants must be used. The search and analysis of such algorithms is left to a future study.

\subsection{Boris Method}
Originally proposed by J. P. Boris during the $4^\text{th}$ International Conference on Numerical Simulation of Plasmas \cite{Boris1971ProceedingsFC}, the Boris pusher was used for the development of a two-dimensional simulation program called CYLRAD that worked with field sources carrying cylindrical symmetry. It is now a classic pusher that is suitable for three dimensions as well and has been adapted for arbitrary order accuracy via an iterative approach \cite{winkel2015borissdc}. Qin \textit{et al.} \cite{qinboris} also discuss some of the many reasons that make it so attractive for both users and researchers. 

Mathematically, the discretized particle push is given by the following set of equations (Ref \cite{Ripperda2017ACC}). 
\begin{eqnarray}
    \label{eqn_boris_average}
    \vec{\bar{v}} &=& (2\Gamma(\vec{u}_e))^{-1}\ \ (\vec{u}_{i+1} + \vec{u}_i)\\
    \label{eqn_boris_pre_update}
    \vec{u}_e &=& \vec{u}_i + \vec{\epsilon}\\
    \label{boris_mag_rotation} \vec{u}_m &=& \vec{u}_e + (\vec{u}_e + (\vec{u}_e \times \vec{\tau})) \times \vec{s}\\
    \vec{u}_{i+1} &=& \vec{u}_m + \vec{\epsilon}
\end{eqnarray}
Here, 
\begin{equation}
    \label{eqn_boris_aux}
    f = \frac{q\delta t}{2m},\ 
    \vec{\tau} = \vec{\beta}/\Gamma(\vec{u}_e),\ 
    \vec{s} = 2\vec{\tau}/(1 + \vec{\tau}^2). 
\end{equation}
The set of equations given by equations \ref{eqn_boris_pre_update}$-$\ref{eqn_boris_aux} correspond to $55$ floating point operations (the vectors $\vec{\epsilon}$ and $\vec{\beta}$ are calculated separately during each iteration, while $f$ is considered constant).

\textbf{Quick note}: Equations \ref{eqn_boris_pre_update}$-$\ref{eqn_boris_aux} corresponds to a simplified form of the original Boris pusher. In his original paper, Boris utilized $\vec{t} = \tan\left({|\vec{\beta}|/\Gamma(\vec{u}_e)}\right)\ \hat{B}$, which yields more accurate results at an increased computing cost. This paper utilizes the simplified form, as is the standard practice in most programs and scientific literature today \cite{birdsall2004plasma}. An alternate form that is only about $1\%-3\%$ more expensive was proposed by Zenitani \textit{et al.}\cite{Zenitani_2018}, who found the original Boris pusher about $46\%$ more expensive than the simplified one.

\subsection{Vay Method}
In a 2007 paper \cite{Vay2007SimulationOB}, Jean-Luc Vay demonstrated that the Boris pusher supports a true solution only in the trivial case when no fields are present. In all other cases, a false force acts on systems undergoing the Boris push, which can introduce deviations from the path that are not born out of true physics. While these deviations aren't significant in the LR regime, they do imply incorrect cancellation of electric and magnetic fields when transforming across frames of reference; something that goes against the very foundation of the theory of relativity. And when the relativistic factor becomes appreciable enough, this can lead to trajectories that diverge significantly from the analytical ones. The Vay integrator has been designed to prevents this spurious force with a more complicated expressions that involve several auxiliary variables for a total cost of 91 operations per step:
\begin{eqnarray}
    \vec{u}_{i+\sfrac{1}{2}} &=& \vec{u}_i + f \left( \vec{E}_{i+\sfrac{1}{2}} + \vec{v}_i \times \vec{B}_{i+\sfrac{1}{2}}  \right)\\
    \vec{u}_e &=& \vec{u}_{i+\sfrac{1}{2}} + \vec{\epsilon}\\
    \vec{u}_{i+1} &=& s \left(\vec{u_e} + (\vec{u}_e \cdot \vec{t})\ \vec{t} + \vec{u}_e \times \vec{t} \right)\\
\end{eqnarray}
Here, the variables involved are 
$u^* = \vec{u}_e \cdot \vec{\beta}/c,\ 
\sigma = (\Gamma(\vec{u}_e))^2 - \vec{\beta}^2,\ 
\vec{t} = \vec{\beta}/\gamma_{i+1},\ 
s = 1 / (1+\vec{t}^2)$ with 
$$\gamma_{i+1} = \sqrt{\frac{\sigma + \sqrt{\sigma + 4 (\vec{\beta}^2 + {u^*}^2) }}{2}}$$

\subsection{Higuera-Cary Method}
The HC integrator was proposed (Ref \cite{Hig2017Integrator}) with the intent to find an integrator that not only gives the best cross-field drift but also preserves phase-space volume; of the Boris and Vay pushers, the latter doesn't conserve volume, as analytically demonstrated by Higuera-Cary. The new integrator proposed utilises the following form of $\vec{\bar{v}}$:
$$\vec{\bar{v}} = (\sfrac{1}{2}\bar{\gamma})^{-1}\ (\vec{u}_i + \vec{u}_{i+1})$$
with
$$\bar{\gamma} = \Gamma\left(\frac{\vec{u}_i + \vec{u}_{i+1}}{2}\right)$$
From thereon, the velocity integration involves the following steps:
\begin{eqnarray}
    \vec{u}_e &=& \vec{u}_i + \vec{\epsilon}\\
    \vec{u}_{m} &=& s \left(\vec{u_e} + (\vec{u}_e \cdot \vec{t})\ \vec{t} + \vec{u}_e \times \vec{t} \right)\\
    \vec{u}_{i+1} &=& \vec{u}_m + \vec{\epsilon} + (\vec{u}_m \times \vec{t})  
\end{eqnarray}
The variables involved in each of these stages have the same definition as the Vay integrator. However, owing to the slight differences involved, the number of operations required is now $88$, as compared to $91$ steps required for Vay integrator. 

\section{Core Times}
\label{times}
This section compares the core run-times of the three particle pushers in the presence of a) only a magnetic field, and b) both electric and magnetic fields. The uniformity and dynamics of the fields themselves are considered irrelevant to computation time. In a comparative analysis, the number of operations required to calculate the fields at different points in space and/or time will be common across all integrators. 

To perform the comparison, we ran our custom code for $600,000$ iterations and dumped the core runtime every $25,000$ iterations. The $\gamma$ update was ignored in the absence of electric fields. To minimize the effects of external factors, each integrator was run 10 times and the average core time plotted. The individual runs have also been shown. Noise in the data was thus easily smoothed out and a linear trend was visible, as is expected. The data being reported is from an Intel(R) Xeon(R) W-1270 CPU @ 3.40GHz processor on an x86 system with 64GB of memory using the gcc version 13.2.0 compiler. However, the relative performance is expected to remain similar for other architectures as well.

\subsection{Only Magnetic Field}
When the system in question consists solely of a uniform magnetic field, the number of operations required at each step is reduced in each of the methods due to the absence of the electric field push. Hence, in an intelligent implementation, changes to $\gamma$, $\vec{t}$, and $\vec{s}$ are not required and this essentially leaves us with only 24 operations for the Boris push. With similar reasoning, Vay and HC integrators require 41 and 38 operations, respectively. Judging solely on the basis of the number of operations required, we would expect the Boris method to be the fastest, while Vay and HC would have similar costs, as already mentioned in ref\cite{Hig2017Integrator}.

\begin{figure}
\centering
    \begin{subfigure}{0.47\textwidth}
        \includegraphics[width=0.9\textwidth]{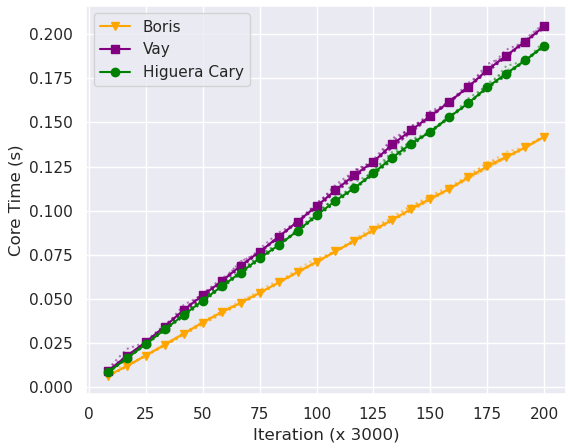}
        \caption{\label{fig_magnetic_core_time}With only magnetic fields.}
    \end{subfigure}
    \begin{subfigure}{0.468\textwidth}
        \includegraphics[width=0.9\textwidth]{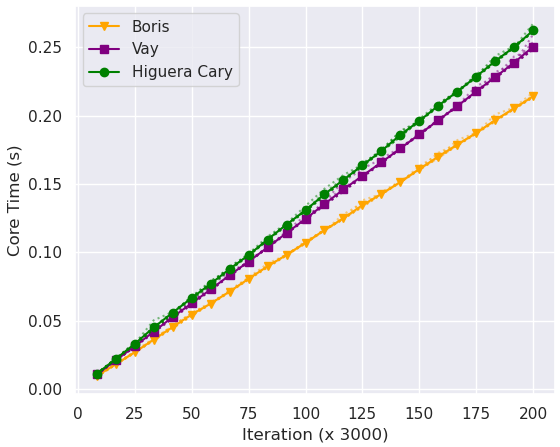}
        \caption{\label{fig_full_core_time}With electric and magnetic fields.}
    \end{subfigure}
    \caption{\label{fig_core_runtimes}Core runtimes (s) of Boris, Vay, and Higuera-Cary integrators. Solid lines correspond to average run-time, while faint dotted lines show the individual runs. The reader is advised to refer to the online version of this text for improved visibility.}
\end{figure}
From figure \ref{fig_magnetic_core_time}, we notice that the Boris pusher performs the most rapid calculations, with $600,000$ iterations taking only about 0.14 seconds. On the other hand, Vay and HC integrators approach the 0.2 second mark. HC integrator is slightly better than Vay, which is found to be the most expensive for the purely magnetic case. 

\subsection{Both Electric and Magnetic Fields}
With both electric and magnetic fields playing a role, we will naturally notice increased times for each of the integrators. The results have been shown in figure \ref{fig_full_core_time}. Boris method increased by about $0.724$ seconds, Vay by $0.462$ seconds, and Higuera-Cary by $0.691$ seconds. This time, the HC integrator is found to be the most expensive, while Boris method once again is the fastest. The Vay and HC integrators in both scenarios have shown similar costs, as also claimed in ref \cite{Hig2017Integrator}. The entirety of these results have been summarised in table \ref{tbl_runtimes}. 
\begin{table}
\centering

    \caption{\label{tbl_runtimes}Core run-times, T,  for Boris, Vay, and HC methods (milliseconds). Average time calculated by slope of first-degree polynomial fit.}
    \begin{tabular}{|l|l|l|l|}
        \hline
        \textbf{Method}& \textbf{System}& \textbf{Avg T/600 steps}& \textbf{Total T (600,000 steps)}\\
        \hline
        \multirow{2}{*}{Boris}& Purely Magnetic& $0.14162$& $141.9397$\\
                              & Electric + Magnetic& $0.21390$& $214.2919$\\
        \hline
        \multirow{2}{*}{Vay}& Purely Magnetic& $0.20398$&$204.3676$\\
                            & Electric + Magnetic& $0.24878$& $250.5632$\\
        \hline
        \multirow{2}{*}{HC}& Purely Magnetic& $0.19303$&$193.4299$\\
                            & Electric + Magnetic& $0.26081$& $262.5085$\\
        \hline
    \end{tabular}
\end{table}
\subsection{What about multi-particle dynamics?}
The above results are for single particles only. However, some perspective about realistic simulations may be gained using the following reasoning:
\begin{itemize}
    \item The typical time-step utilized for fine-grained plasma simulations is around $10^{-2}\ T_c$, i.e., about 100 iterations are needed to complete one gyration period. For systems with only slowly varying processes, this can be increased to $0.1\ T_c$, but we err on the side of caution.
    \item This would imply that to resolve 6 complete gyrations, we would require 600 iterations. This corresponds to $\approx$ 0.15 ms for Boris, and 0.2 ms for Vay and HC integrators in the purely magnetic case.
    \item Plasma density can vary anywhere between $10^6$ to $10^{28}\ m^{-3}$, as stated in ref \cite{chen2012introduction}. Hence, a $1\ \text{mm}^3$ system would contain a maximum of $10^{19}$ particles. However, the $N^2$ operations can be significantly brought down using PIC simulations \cite{birdsallPIC}. Let us suppose that this gets us to around $10^6$ operations per iteration.
    \item Using the above information, we can conclude that such a system would require no more than $150-200$ seconds ($\approx 2.5-3.3$ minutes) for the \textbf{temporal push}. Naturally, this is a highly artificial number since actual plasma simulations would also involve spatial push as well as Maxwell equation solution, which will significantly increase the total core time.
\end{itemize}

\section{Performance}
We now present the results for each of the integrators in terms of accuracy. The LR, HR, and UR regimes have been studied in the following sections. In each of these regimes, we have studied the magnetic gyration and cross-field drift of a positron. The choice between a positron and an electron is entirely arbitrary and a matter of convenience. Change to the sign of the charge will not affect the accuracy of the outcome. Besides, positron acceleration via plasmas is anyway a lucrative field \cite{cao2024plasmapositron} and thus, we are certain that this choice is not ill-suited.

For each of the cases, we utilize a coarse time-grid, typically around $\sfrac{T_C}{20}$, with minor adjustments to make the calculations and analysis tractable. With a fine grid, the errors would have been significantly lower, but a much larger number of iterations would have been required to resolve the errors. The magnetic field is applied in the Z-direction with magnitudes of $5$ T and $10^{-4}$ T for gyration and cross field cases, respectively. The electric field is $100\ \text{V}\cdot\text{m}^{-1}$ in the X-direction. This means the cross field drift would be $10^6\ \text{m}\cdot\text{s}^{-1}$ in the negative Y-direction. Various quantities have been compared and the results follow.

\subsection{Low Relativistic Regime}
\label{low_rel}
\subsubsection{Magnetic Gyration}
We initialized a positron with $\gamma = 1.27282$, and $\omega = 6.90 \times 10^{11}\ \text{rad.s}^{-1}$ and studied it using a time grid of $dt \approx \sfrac{T_c}{18}$ over approximately 8 gyration periods (150 iterations). 

\begin{figure}
\centering
\begin{subfigure}{0.45\textwidth}
    \includegraphics[width=0.9\textwidth]{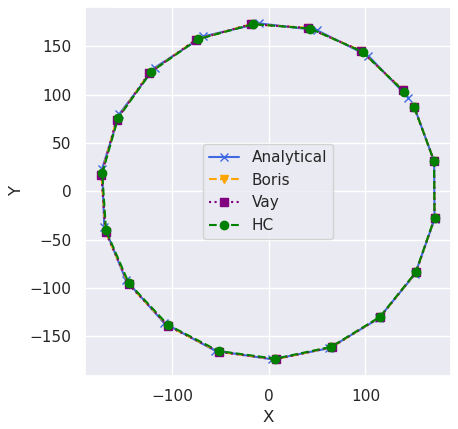}
    \caption{\label{fig_lrg_config_start}The first cycle.}
\end{subfigure}
\begin{subfigure}{0.45\textwidth}
    \includegraphics[width=0.9\textwidth]{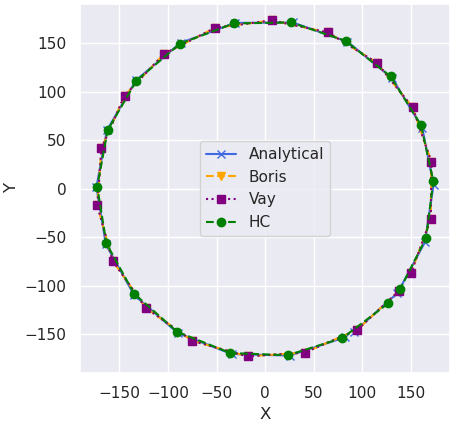}
    \caption{\label{fig_lrg_config_last}The last cycle.}
\end{subfigure}
\caption{\label{fig_config_lrg}Configuration space plots in microns for the uniform magnetic field in the LR regime.}
\end{figure}

In the presence of solely a magnetic field, the particle is expected to execute a helical trajectory.  Since motion along the z-direction is trivial, we plot the first cycle on the X-Y configuration space in figure \ref{fig_lrg_config_start} for the first gyration period. While the three integrators' overall plots merge seamlessly with the analytical trajectory, we would like to highlight the phase errors that have evolved. It may be noticed that in the beginning, the markers almost completely overlap but they draw further apart as the particle progresses, until during the 8$^{th}$ oscillation (figure \ref{fig_lrg_config_last}), the discrepancy is vastly pronounced.

\begin{figure}
\centering

    \begin{subfigure}{0.51\textwidth}
        \includegraphics[width=0.9\textwidth]{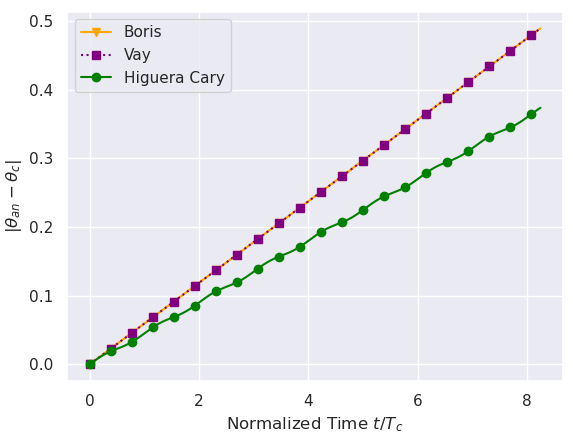}
        \caption{\label{fig_lrg_phase}Phase error.}
    \end{subfigure}
    \begin{subfigure}{0.475\textwidth}
        \includegraphics[width=0.9\textwidth]{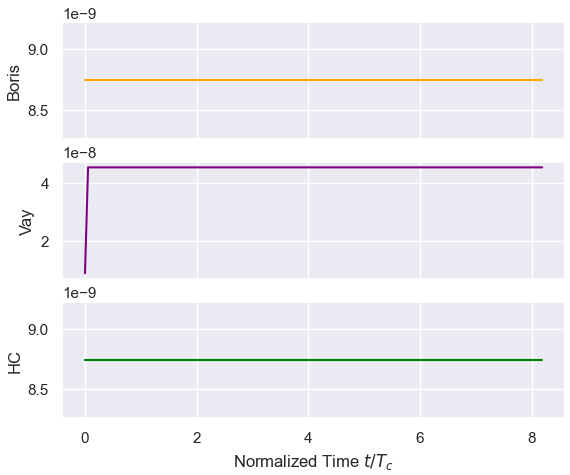}
        \caption{\label{fig_lrg_gamma}Gamma error.}
    \end{subfigure}
    
\caption{\label{fig_lrg_phase_plus_gamma}Relative Gamma and absolute phase errors (rad) for magnetic gyration in the LR regime. The slight wave in HC phase error stems out of the sinusoidal nature of the trajectory.}
\end{figure}

Speaking of phase errors, we also take the liberty of plotting them in figure \ref{fig_lrg_phase}. As can be noticed via the configuration space as well, while the Boris and Vay methods introduce similar amounts of phase lag, the HC method is significantly better. The phase lag grows linearly for each of the methods and by the $8^\text{th}$ cycle, has gone up to $0.374\ \text{rad}$ for the HC method, and $0.489\ \text{rad}$ for Boris and Vay methods. This might appear significant, but we would like to remind that this run is using a very coarse grid of $\sfrac{T_c}{18}$. With a finer grid, the results will improve. It should also be noted that for clarity, not all data points have been shown with markers in the graph. Also, the phase errors appear to carry a slight wave. This is because the trajectory is sinusoidal and we have used the \texttt{numpy.unwrap} functionality to analyze the data

\begin{figure}
\centering

\begin{subfigure}{0.468\textwidth}
    \includegraphics[width=0.9\textwidth]{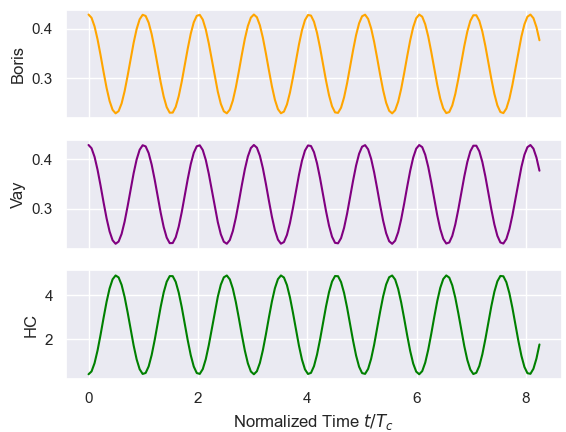}
    \caption{\label{fig_lrg_radius}Relative errors.}
\end{subfigure}
\begin{subfigure}{0.49\textwidth}
    \includegraphics[width=0.9\textwidth]{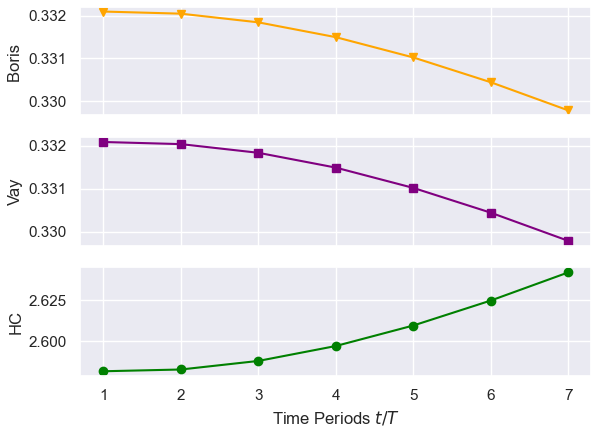}
    \caption{\label{fig_lrg_radius_mean}Mean of relative errors.}
\end{subfigure}
\caption{\label{fig_lrg_radius_errors}Gyroradius errors ($\times 10^{-3}$) for the uniform magnetic field in the LR regime.}
\end{figure}

Furthermore, while it is not evident from the configuration space plot, there are minor errors in the gyroradius for each of the three cases, which are plotted in figures \ref{fig_lrg_radius} and \ref{fig_lrg_radius_mean}. The error appears to have a cyclic nature, which we chalk up to the way the trajectory is estimated by each of the methods. The simulated trajectory essentially zigzags around the analytical one, and we have plotted the mean of the errors to get a better idea. Boris and Vay methods develop similar order errors ($0.057\ \mu m$ for an actual gyroradius of $173.74\ \mu m$), with Vay method proving slightly better off. HC method is an order of magnitude worse ($\approx 0.45\ \mu m$ of error). Note that while the radius appears to be cyclic, the mean isn't exactly constant since it is taken over the time period of particle gyration, which may not coincide with the time period of the error itself.

Finally, we plot the error in the relativistic factor in figure \ref{fig_lrg_gamma}, which is expected to be minute and constant since no energy is being imparted to the particles. Each integrator shows the same trend, save the Vay integrator, which takes a few iterations to settle into its steady value. The Vay error is one order of magnitude higher but overall, the error itself is too small to worry about ($10^{-9}-10^{-8}$). Boris and HC methods have almost the same performance.

\subsubsection{Cross Field Drift}
Under the action of crossed electric and magnetic fields, charged particles execute magnetic gyration along with additional drift in a direction perpendicular to both the fields. Its magnitude is given by $|\vec{v}_D|=\sfrac{\vec{E}}{\vec{B}}$ and it is the cause of a large number of issues and instabilities in plasma physics \cite{chang1990instability}. 

There are two ways to study the cross-field drift. The analytical solution becomes significantly simpler if one shifts to a frame of reference moving with a velocity $\vec{v}_D$ with respect to the lab frame. In this frame, the particle merely gyrates at a reduced frequency and does not drift. It is also possible to analyze the particle in the lab frame itself. For cases where acceleration does not significantly affect the gyration frequency, the latter analysis suffices. However, the former method is more tractable otherwise and we will opt for that when studying the UR regime. 

\begin{figure}
\centering

\begin{subfigure}{0.4\textwidth}
    \includegraphics[width=\linewidth]{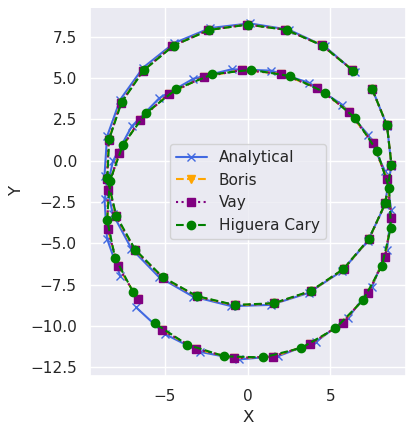}
    \caption{\label{fig_lrd_config} Configuration Space}
\end{subfigure}
\begin{subfigure}{0.58\textwidth}
    \includegraphics[width=\linewidth]{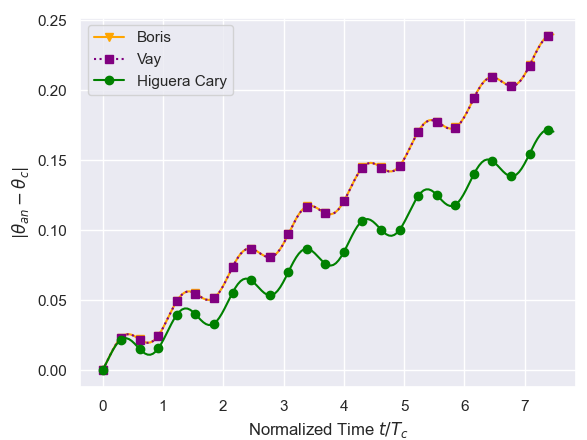}
    \caption{\label{fig_lrd_phase}Phase errors.}
\end{subfigure}
\caption{\label{fig_lrd_config_plus_phase}Configuration space in metres and absolute phase errors (rad) for cross-field drift in the LR regime.}
\end{figure}

We initialized a positron just as before and study it with time-step $dt \approx \sfrac{T_C}{22} = 2 \times 10^{-8}\ \text{s}$. The first and last $22$ iterations are plotted in figure \ref{fig_lrd_config}. The drift in the y-direction is clearly visible, and we again notice phase errors that have crept up in the trajectory. These are plotted in figure \ref{fig_lrd_phase}. We notice that the HC method has again emerged as a better candidate, going up to a maximum of $0.17\ \text{rad}$ over the 7 cycles. On the other hand, Boris and Vay methods develop a phase lag of $0.24\ \text{rad}$ each.

\begin{figure}
\centering

    \includegraphics[width=0.6\linewidth]{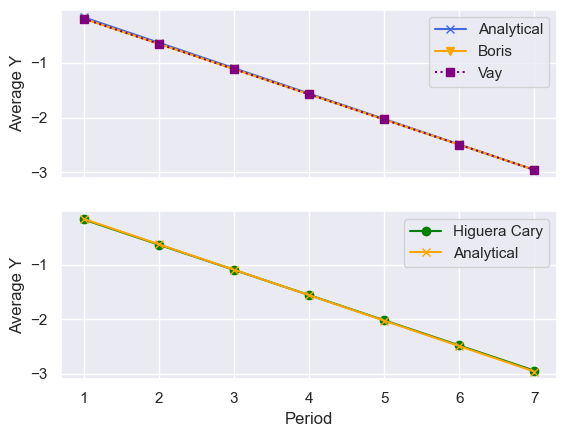}
    \caption{\label{fig_lrd_y_mean}Average y over time ($\left<y\right>$) for cross-field drift in the LR regime (metres).}
\end{figure}

To study the drift, we have plotted average $\left<y\right>$ in figure \ref{fig_lrd_y_mean}. The overlap of the simulated average with the actual average shows us that the cross-field drift has been captured well. Further, the slope of the average for each of the integrators has been compared with the analytical average, allowing us to determine just how well the drift motion is captured. This is summarised in table \ref{tbl_lrd_y_drift}. We notice that the Vay integrator is very slightly better in terms of capturing cross-field drift. The difference isn't that pronounced when $\gamma$ is so low, but it is there. The performance of HC in terms of cross-field drift is slightly worse but still very much acceptable.

\begin{table}
\centering

    \caption{\label{tbl_lrd_y_drift}Slopes of $\left<y\right>$ for each of the integrators.}
    \begin{tabular}{|l|l|l|l|}
        \hline
        \textbf{Method} & \textbf{Slope} & \textbf{Absolute Error} & \textbf{Percentage Error}\\
        \hline
        Analytical \phantom{\quad} & $-0.46631098$ \phantom{\qquad} & -- & --\\
        \hline
        Boris & $-0.46191718$ & $0.00439380$ \phantom{\qquad} & $0.942\%$ \phantom{\qquad}\\
        \hline
        Vay & $-0.46191934$ & $0.00439164$ \phantom{\qquad} & $0.941\%$ \phantom{\qquad}\\
        \hline
        HC & $-0.46178848$ & $0.00452251$ \phantom{\qquad} & $0.969\%$ \phantom{\qquad}\\
        \hline
    \end{tabular}
\end{table}

\subsection{High Relativistic Regime}
\label{high_rel}

\subsubsection{Magnetic Gyration}
To study the HR regime, we initialized a positron with $\gamma = 3.589$ and $T_C = 2.56545 \times 10^{-11}\ \text{s}$. We used a temporal grid with size $dt \approx \sfrac{T_C}{25} = 10^{-12}\ \text{s}$, and resolved 6 gyrations across 170 steps.

\begin{figure}

\begin{subfigure}{0.47\textwidth}
\centering

    \includegraphics[width=0.9\textwidth]{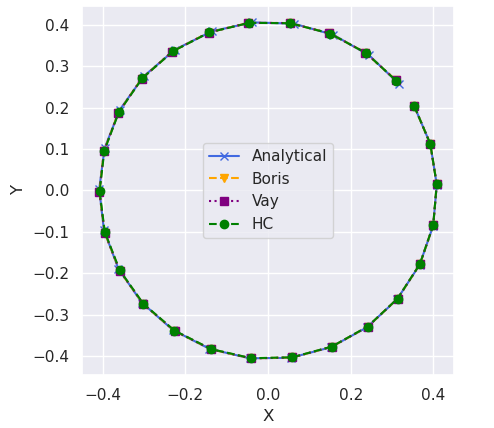}
    \caption{\label{fig_hrg_config_start}The first cycle.}
\end{subfigure}
\begin{subfigure}{0.47\textwidth}
    \includegraphics[width=0.9\textwidth]{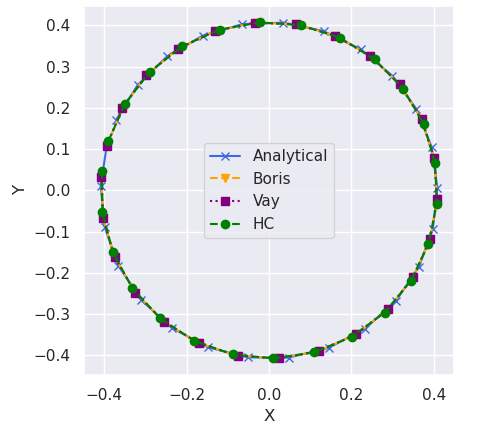}
    \caption{\label{fig_hrg_config_last}The last cycle.}
\end{subfigure}
\caption{\label{fig_config_hrg}Configuration space plot in mm for uniform magnetic field in the HR regime.}
\end{figure}

\begin{figure}

    \begin{subfigure}{0.49\textwidth}
        \includegraphics[width=0.9\textwidth]{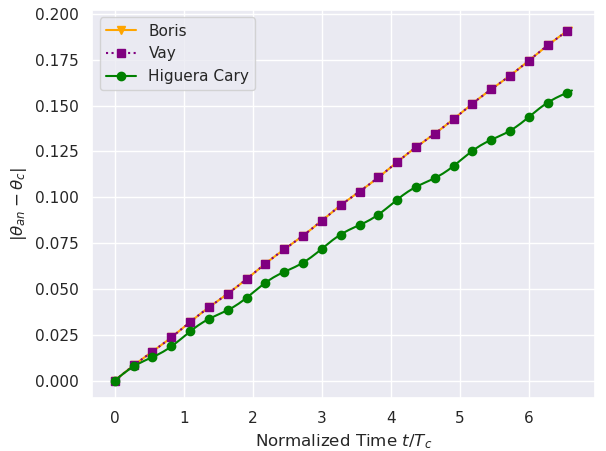}
        \caption{\label{fig_hrg_phase}Phase error.}
    \end{subfigure}
    \begin{subfigure}{0.48\textwidth}
        \includegraphics[width=0.9\textwidth]{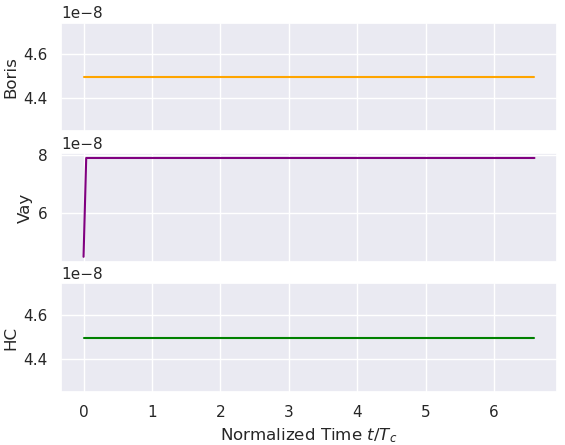}
        \caption{\label{fig_hrg_gamma}Gamma error.}
    \end{subfigure}
\caption{\label{fig_hrg_phase_plus_gamma}Relative Gamma error and absolute phase errors (rad) for magnetic gyration in the HR regime.}
\end{figure}

Once again, the simulated trajectory matches the analytical one (see figures \ref{fig_hrg_config_start}, \ref{fig_hrg_config_last}, which plot the first and last $26$ steps on the configuration space), albeit with some phase error, which is evident in the markers in figure \ref{fig_hrg_config_last}. The phase error is plotted in figure \ref{fig_hrg_phase}. As before, Boris and Vay methods develop almost the same phase error ($\approx 0.1926\ \text{rad}$), while HC method develops $0.1583\ \text{rad}$ over more than 6 cycles. This is better than the LR case discussed in section \ref{low_rel}, but we also note that the step size is a little finer this time around and thus, it is to be expected.

\begin{figure}
\centering

\begin{subfigure}{0.49\textwidth}
    \includegraphics[width=0.9\textwidth]{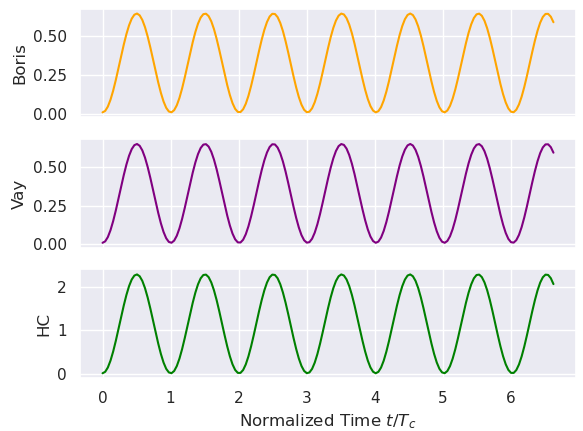}
    \caption{\label{fig_hrg_radius}Relative errors ($\times 10^{-3}$).}
\end{subfigure}
\begin{subfigure}{0.49\textwidth}
    \includegraphics[width=0.9\textwidth]{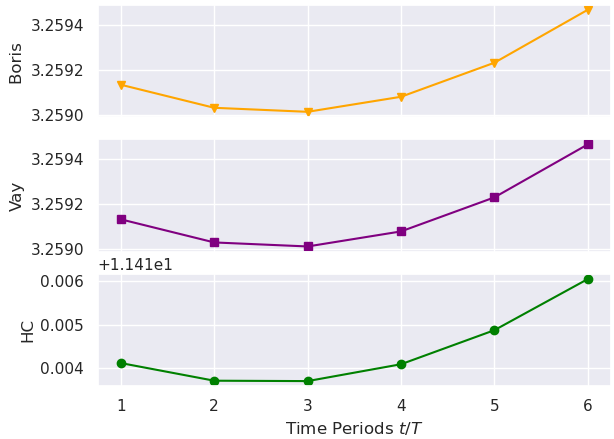}
    \caption{\label{fig_hrg_radius_mean}Mean of relative errors ($\times 10^{-4}$).}
\end{subfigure}
\caption{\label{fig_hrg_radius_errors}Gyroradius errors for the uniform magnetic field in the HR regime.}
\end{figure}

The relative errors in gyroradius and their means are plotted in figures \ref{fig_hrg_radius}, and \ref{fig_hrg_radius_mean}, respectively. The theoretical radius is $0.4083\ \text{mm}$, against which, Boris and Vay methods develop relative errors of the order of $0.0325\%$, while the relative error in HC method is of the order of $0.1141\%$. 

Finally, the error in $\gamma$ is plotted in figure \ref{fig_hrg_gamma}. The error remains constant for each method, though Vay error takes a few iterations to settle into a higher error value. Overall, the relative error is only of the order of $10^{-6}\%$ and thus, negligible.

\subsubsection{Cross Field Drift}
Cross field drift was studied using $\vec{B} = 10^{-4}\ \hat{k}$ and $\vec{E} = 100\ \hat{i}$. The time period of oscillation and the step-size were $T_C = 1.287 \times 10^{-6}\ \text{s}$ and $dt \approx \sfrac{T_C}{22} = 6 \times 10^{-8}\ \text{s}$. The program was run for 180 steps, corresponding to almost 9 oscillation cycles.

\begin{figure}
\centering

\begin{subfigure}{0.39\textwidth}
    \includegraphics[width=\linewidth]{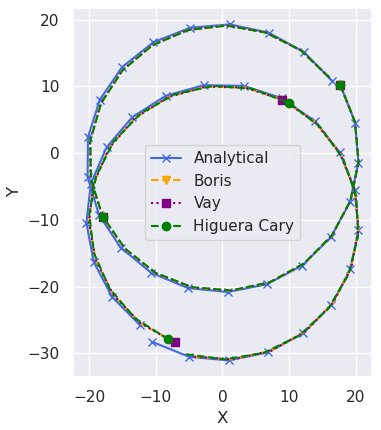}
    \caption{\label{fig_hrd_config}Configuration space.}
\end{subfigure}
\begin{subfigure}{0.6\textwidth}
    \includegraphics[width=\linewidth]{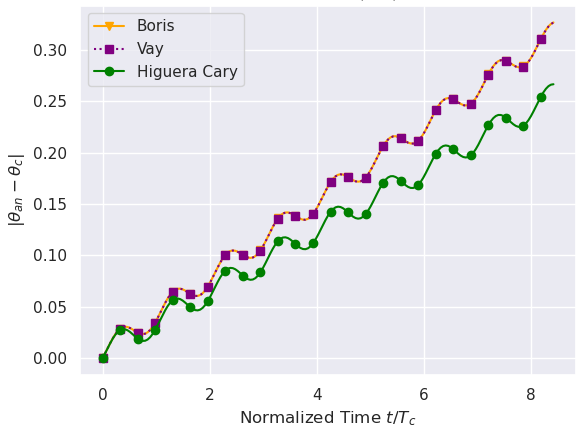}
    \caption{\label{fig_hrd_phase}Phase errors.}
\end{subfigure}
\caption{\label{fig_hrd_config_plus_phase}Configuration space (metres) and absolute phase errors (rad) for cross-field drift in the HR regime.}
\end{figure}

As before, the first and last cycle on the configuration space is plotted in figure \ref{fig_hrd_config}. The drift is visible, as is the phase error, which is plotted in figure \ref{fig_hrd_phase}. It can be seen that the HC method carries less phase error. It develops a phase lag only of the order of $0.2663\ \text{rad}$ over the course of 8 cycles, whereas Boris and Vay integrators develop similar phase lags of the order of $0.3264\ \text{rad}$ and $0.3263\ \text{rad}$, respectively.

\begin{figure}
\centering
    
    \includegraphics[width=0.6\linewidth]{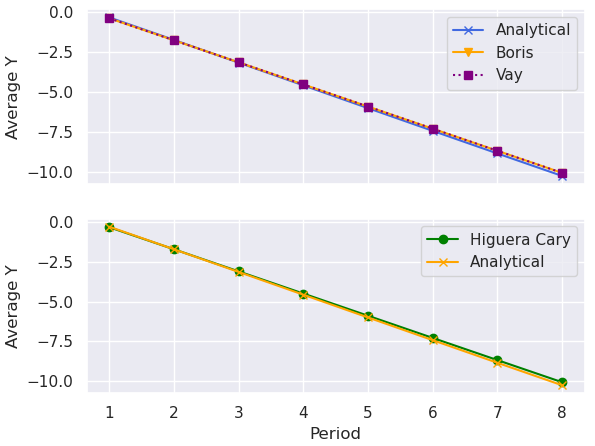}
    \caption{\label{fig_hrd_y_mean}Average y over time ($\left<y\right>$) for cross-field drift in the HR regime (metres).}
\end{figure}

Cross-field drift is visible in the phase space (figure \ref{fig_hrd_config}), and we again compare the slope of average y for each of the integrators. This is plotted in figure \ref{fig_hrd_y_mean} and tabulated in table \ref{tbl_hrd_y_drift}. 

\begin{table}
\centering

    \caption{\label{tbl_hrd_y_drift}Slopes of $\left<y\right>$ for each of the integrators in the HR regime. The slopes are more negative now since the gyroradius has changed.}
    \begin{tabular}{|l|l|l|l|}
        \hline
        \textbf{Method} & \textbf{Slope} & \textbf{Absolute Error} & \textbf{Percentage Error}\\
        \hline
        Analytical \phantom{\quad} & $-1.41998544$ \phantom{\qquad} & -- & --\\
        \hline
        Boris & $-1.38362530$ & $0.03636014$ \phantom{\qquad}& $2.561\%$ \phantom{\qquad}\\
        \hline
        Vay & $-1.38363835$ & $0.03634709$ \phantom{\qquad} & $2.560\%$ \phantom{\qquad}\\
        \hline
        HC & $-1.38729913$ & $0.03268632$ \phantom{\qquad} & $2.302\%$ \phantom{\qquad}\\
        \hline
    \end{tabular}
\end{table}

\subsection{Ultra Relativistic Regime}
\label{ultra_rel}

\subsubsection{Magnetic Gyration}
The UR regime presents the most interest to the plasma community and previous works undertaken also target this region predominantly (ref \cite{Ripperda2017ACC,Hig2017Integrator,Vay2007SimulationOB}). We present the findings for a positron initialized with $\gamma = 56.95$ and $T_C = 4.07 \times 10^{-10}\ \text{s}$ by running the program for 180 steps with a temporal grid of $dt \approx \sfrac{T_C}{20} = 2 \times 10^{-11}\ \text{s}$.

\begin{figure}
\centering

\begin{subfigure}{0.47\textwidth}
    \includegraphics[width=0.9\textwidth]{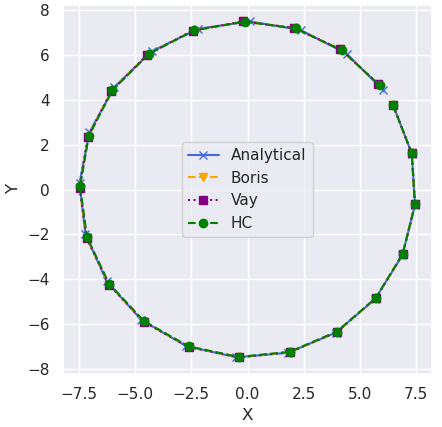}
    \caption{\label{fig_urg_config_start}The first cycle.}
\end{subfigure}
\begin{subfigure}{0.47\textwidth}
    \includegraphics[width=0.9\textwidth]{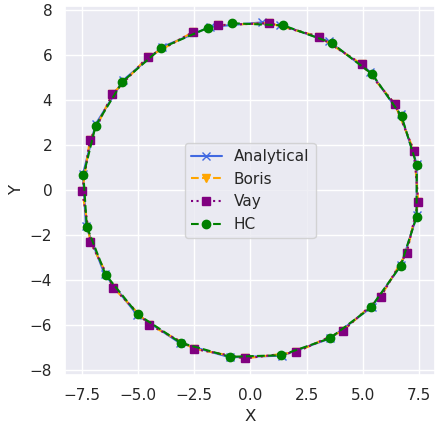}
    \caption{\label{fig_urg_config_last}The last cycle.}
\end{subfigure}
\caption{\label{fig_config_urg}Configuration space plot (mm) for uniform magnetic field in the UR regime.}
\end{figure}

\begin{figure}
\centering

    \begin{subfigure}{0.485\textwidth}
        \includegraphics[width=0.9\textwidth]{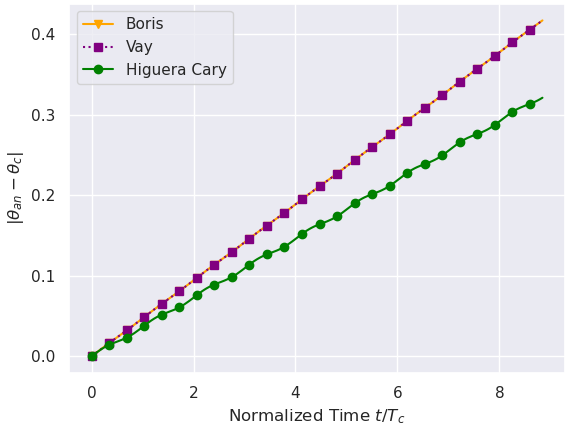}
        \caption{\label{fig_urg_phase}Phase error.}
    \end{subfigure}
    \begin{subfigure}{0.481\textwidth}
        \includegraphics[width=0.9\textwidth]{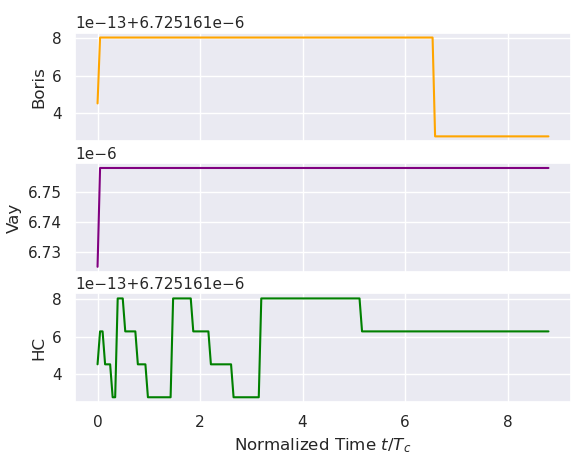}
        \caption{\label{fig_urg_gamma}Gamma error.}
    \end{subfigure}
\caption{\label{fig_urg_phase_plus_gamma}Relative Gamma error and absolute phase errors (rad) for magnetic gyration in the UR regime.}
\end{figure}

The configuration space plots, showing the first and last 21 iterations are shown in figures \ref{fig_urg_config_start} and \ref{fig_urg_config_last}, respectively. The trajectory matches well with the analytical value, and as before, pronounced phase errors are visible for the Boris and Vay integrators. These are plotted in figure \ref{fig_urg_phase}. 

The discrepancy in phase errors between the HC method and the other two methods is more significant this time around. HC method develops a phase lag of the order of $0.3213\ \text{rad}$, compared to the Boris and Vay integrators, which have phase lags of $0.4173\ \text{rad}$ each.

\begin{figure}
\centering

\begin{subfigure}{0.49\textwidth}
    \includegraphics[width=0.9\textwidth]{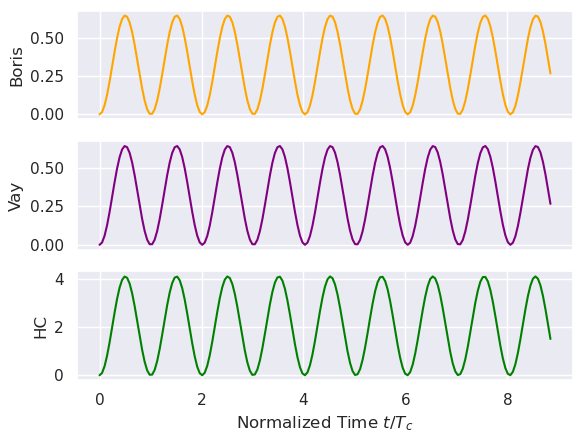}
    \caption{\label{fig_urg_radius}Relative errors ($\times 10^{-3}$).}
\end{subfigure}
\begin{subfigure}{0.49\textwidth}
    \includegraphics[width=0.9\textwidth]{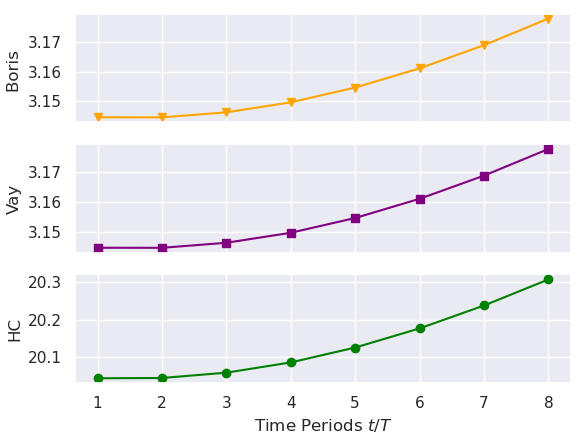}
    \caption{\label{fig_urg_radius_mean}Mean of relative errors ($\times 10^{-4}$).}
\end{subfigure}
\caption{\label{fig_urg_radius_errors}Gyroradius errors for the uniform magnetic field in the UR regime.}
\end{figure}

The relative errors in gyroradius and their means are plotted in figures \ref{fig_urg_radius} and \ref{fig_urg_radius_mean}, respectively. Against a true value of $7.4826\ \text{mm}$, Boris and Vay errors yield absolute errors of the order of $0.00235\ \text{mm}$, while HC error is an order higher at $0.015\ \text{mm}$.

The error in $\gamma$ yields interesting behaviour in the UR regime (figure \ref{fig_urg_gamma}). The Boris and Vay errors are no longer constant and we start seeing staggered patterns, which are more pronounced in HC method. The Vay integrator shows the same behaviour as before.

\subsubsection{Cross Field Drift}
As previously stated, cross-field drift this time around is being studied in the frame moving with the cross-field drift wherein, the particle experiences a vanishing electric field and a reduced magnetic field of $\vec{B}' = \vec{B}/\kappa$ and thus, performs a purely gyrating motion. here, $\kappa$ is the relativistic factor associated with the drift motion of the particle, i.e., $\kappa = 1/\sqrt{1 - |\vec{v}_D|^2/c^2}$. In the UR case, changes to $\gamma$ affect the movement of the particle and thus, studying performance with an electric field present is difficult. Therefore, we have chosen to perform our analysis in the shifted reference frame. However, the simulation is performed in the laboratory frame only.

To that end, a positron was initialized with $|\vec{v_\perp}| = 0.384c$, $v_{||} = 0.923c$ such that $T_C = 6.69 \times 10^{-06}\ \text{s}$, which was covered using a time-step of $dt\approx \sfrac{T_C}{23} = 3 \times 10^{-7}\ \text{s}$ for 8 cycles of motion. 

In the shifted coordinates, we now have $\kappa = 1.002$, $\vec{B}'=9.9833 \times 10^{-5}\ \hat{j}$, and $\vec{E} = \vec{0}$. The gyration of the particle is now at a frequency determined by the reduced magnetic field $\omega' = \sfrac{qc|\vec{B}'|} {|\vec{p'}|} $, with the cyclotron radius given by $r' = \sfrac{|\vec{p}'_\perp|} {q|\vec{B}'|}$.

\begin{figure}
\centering

\begin{subfigure}{0.47\textwidth}
    \includegraphics[width=0.9\textwidth]{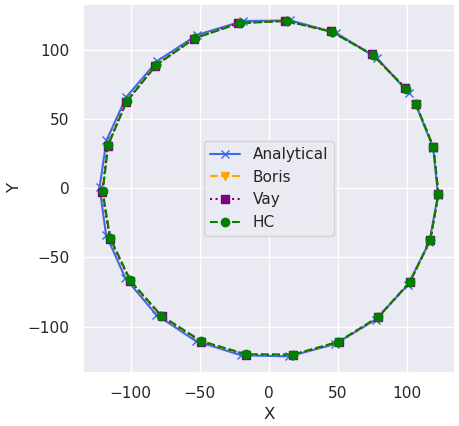}
    \caption{\label{fig_urd_config_start}The first cycle.}
\end{subfigure}
\begin{subfigure}{0.47\textwidth}
    \includegraphics[width=0.9\textwidth]{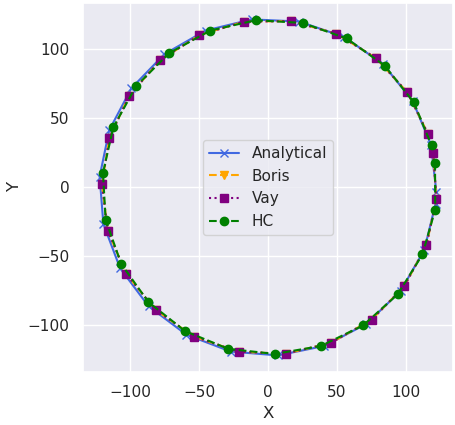}
    \caption{\label{fig_urd_config_last}The last cycle.}
\end{subfigure}
\caption{\label{fig_config_urd}Configuration space plot (metres) for cross field drift plotted in the shifted frame in UR regime.}
\end{figure}

\begin{figure}
\centering

    \begin{subfigure}{0.48\textwidth}
        \includegraphics[width=0.9\textwidth]{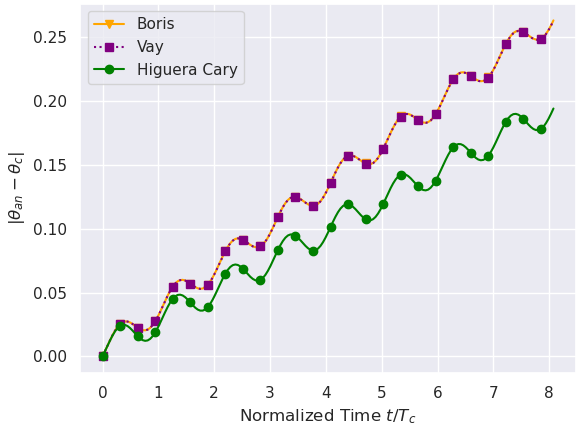}
        \caption{\label{fig_urd_phase}Phase error.}
    \end{subfigure}
    \begin{subfigure}{0.5\textwidth}
        \includegraphics[width=0.9\textwidth]{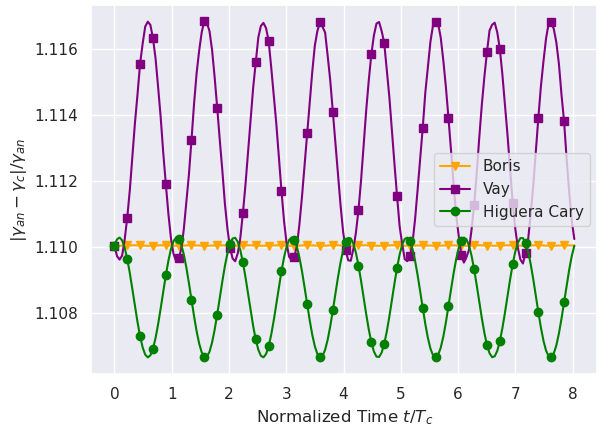}
        \caption{\label{fig_urd_gamma}Gamma error.}
    \end{subfigure}
\caption{\label{fig_urd_phase_plus_gamma}Relative gamma error and absolute phase errors (rad) for cross field drift plotted in the shifted frame in UR regime.}
\end{figure}

The configuration space plot, with the first and last 23 steps, is shown in figures \ref{fig_urd_config_start} and \ref{fig_urd_config_last}, respectively. The phase errors developed are plotted alongside in figure \ref{fig_urd_phase}. Boris and Vay methods develop higher errors of $0.263\ \text{rad}$, while HC method fares better at $0.194\ \text{rad}$.

Figure \ref{fig_urd_gamma} also shows the error in relativistic factor for each of the methods. Note that this error is in the relativistic factor that the methods develop in the boosted frame where no electric field is present. As is visible, the performance is remarkable, with percentage errors only in the range of $0.1\%$. The HC integrator fares either better or the same as the Boris push, while the Vay integrator has a much more unstable nature.

\begin{figure}
\centering

\begin{subfigure}{0.485\textwidth}
    \includegraphics[width=0.9\textwidth]{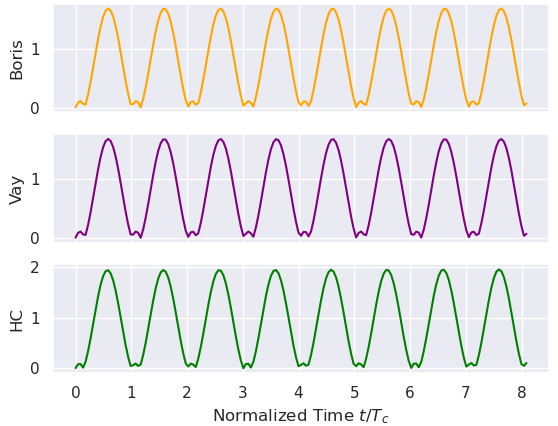}
    \caption{\label{fig_urd_radius}Relative errors ($\times 10^{-2}$).}
\end{subfigure}
\begin{subfigure}{0.495\textwidth}
    \includegraphics[width=0.9\textwidth]{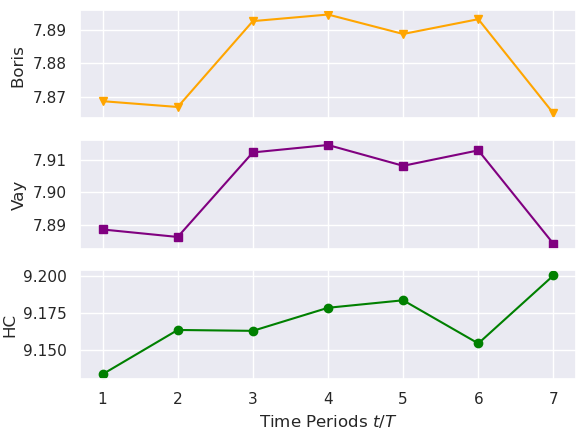}
    \caption{\label{fig_urd_radius_mean}Mean of relative errors ($\times 10^{-3}$).}
\end{subfigure}
\caption{\label{fig_urd_radius_errors}Gyroradius errors for cross field drift plotted in the shifted frame in UR regime.}
\end{figure}

Finally, the relative errors in gyroradius and their means in the boosted frame are plotted in figures \ref{fig_urd_radius} and \ref{fig_urd_radius_mean}, respectively. The performance is comparable for all integrators, but the HC integrator is slightly worse off (average absolute error of $1.1002$ as compared to $0.937$ and $0.939$ for Boris and Vay pushes). The relative errors are also slightly increased for the Vay integrator than the Boris push ($0.937$ absolute error compared to $0.939$).  

\section{The Fitness Parameter}
\label{accuracy_param}
The final aim of this text is to propose a parameter that we believe, can serve as a one-stop shop for confirming whether a certain integrator is suitable for a given task. After studying the three methods we've concerned ourselves within this text, we are now in a position to complete this goal. A comprehensive parameter must, in principle: a) factor in the computational cost, b) factor in the error accumulated, and, c) factor in the response of the method to a particular set of conditions. Keeping these factors in mind, we propose a "Fitness Parameter" as 
$$f \equiv \frac{1}{\kappa}e^{-\epsilon}$$
where $\kappa=$ computational cost per iteration (FLOPs), and $\epsilon=\log_{10}$ of relative error developed across five time periods of the slowest process in the system, using a time-step of $\sfrac{T}{50}$. $T$ is the time period of the fastest process in the system. This way, the time-step is fine enough to resolve the fast-moving processes, and the integrator goes through enough iterations to resolve the slow processes as well. If the relative error oscillates, it is recommended to take the per-period average of relative error and use the average in the fifth period. Two important points to note are:
\begin{enumerate}
    \item The computational cost can change depending on the context and implementation. It is proposed that the cheapest implementation for a given context be used to calculate $f$.
    \item As to the question of which quantity's relative error to use for calculating $f$, we propose that the energy be used as the quantity of interest. The reason is simple: energy conservation implies, in almost all cases, a perfect trajectory calculation. Any numerical errors will necessarily give rise to energy errors and thus, $f$ will incorporate it all.
\end{enumerate}

\begin{figure}
\centering

\includegraphics[width=0.9\textwidth]{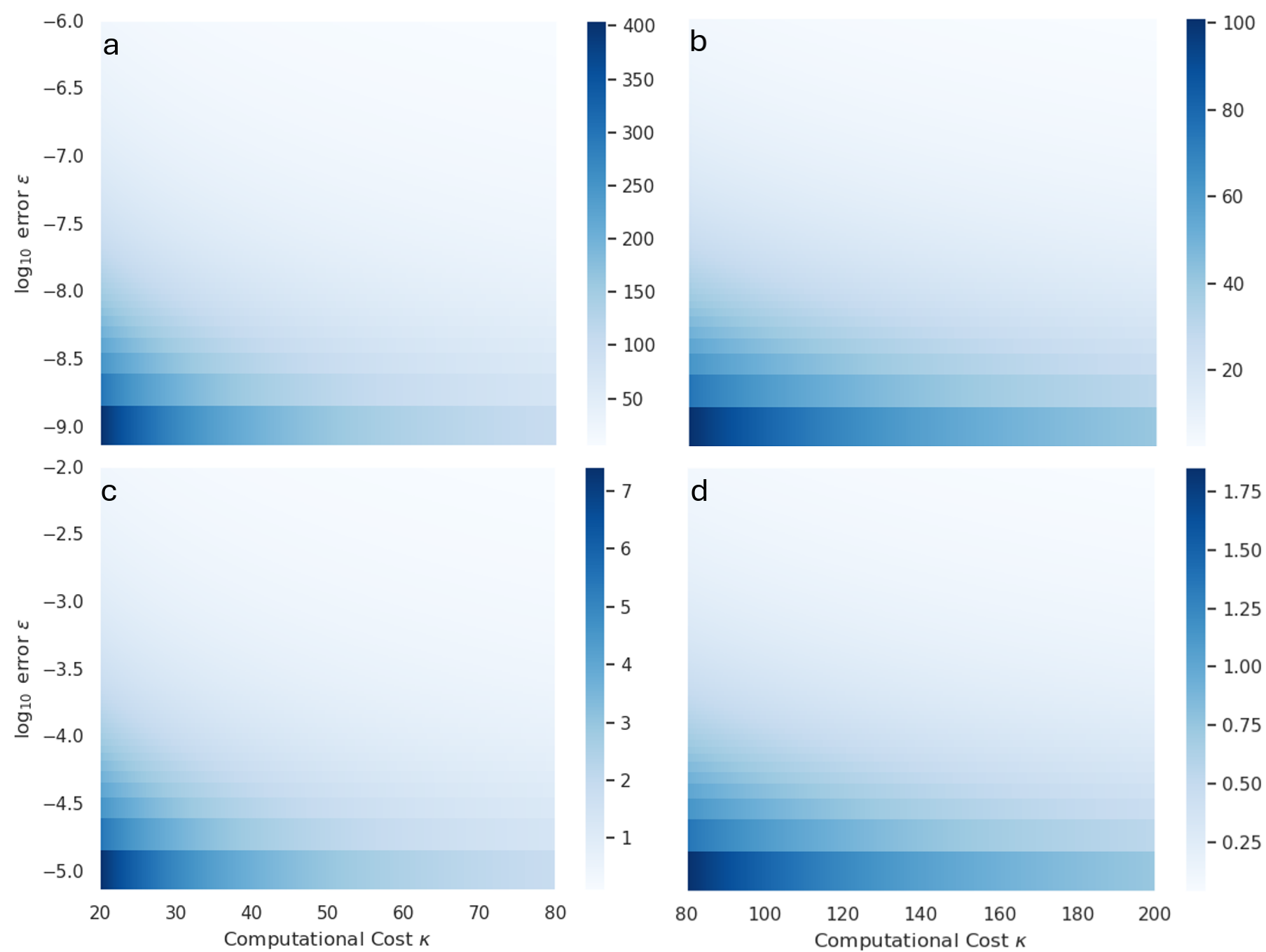}
\caption{\label{fig_f_params}Variation of fitness parameter $f$ with (a) low $\kappa$ low $\epsilon$, (b) high $\kappa$ low $\epsilon$, (c) low $\kappa$, high $\epsilon$, and (d) high $\kappa$ high $\epsilon$.}
\end{figure}
By definition, $f$ conveys information about the computational cost of an integrator while also factoring in the error via $\epsilon$. $f$ will decrease with $\kappa$ and increase with $\epsilon$ (figure \ref{fig_f_params}) and thus, a higher $f$ will imply a more suitable integrator than one with lower $f$. Also, as is clear from the figure, the absolute value of the $f$ parameter is irrelevant. What matters is how far apart two integrators are in a particular context.

The fitness parameters for the Boris, Vay, and HC methods have been tabulated in table \ref{tbl_fitness_param}. It must be noted that for an optimized implementation, the computational cost must change when either the magnetic or the electric field is absent, and thus, $f$ will also change.
\begin{table}
\centering

\caption{\label{tbl_fitness_param}Fitness Parameters for the Boris, Vay, and HC methods.}
    \begin{tabular}{|l|l|l|l|l|}
        \hline
        \multirow{2}{*}{\textbf{Method}}  & \multicolumn{2}{c|}{\textbf{Only Magnetic Field}} & \multicolumn{2}{c|}{\textbf{Cross $\vec{E}$ and $\vec{B}$ Fields}}\\
        \cline{2-5}& Cost (FLOPs) & f parameter& Cost (FLOPs) & f parameter\\
        \hline
        Boris &  24&  68.3469& 55 &0.0308\\
        Vay &  41&  19.9109& 91 &0.0186\\
        Higuera-Cary &  38&  43.1664& 88&0.0192\\
        \hline
    \end{tabular}
\end{table}

These parameters are consistent with the timings and results we have found so far. In the absence of an electric field, the Vay pusher is slightly more expensive, and thus, even though the errors are a little higher, its $f$ parameter is a bit lower. The Boris pusher, being almost twice as efficient has a much higher $f$ parameter. Similarly, in the cross-field case, we have found that the Boris pusher is highly efficient and the errors are comparable, leading to a much higher $f$ parameter. On the other hand, the HC pusher, has a higher cost, and thus, $f$ is significantly lower.

\section{Summary}
\label{conclusion}
A study has been performed of the low-, high-, and UR performances of the Boris, Vay, and HC pushers using a custom code titled \textbf{PaTriC}. References and reviews are made of previous works and similar outcomes have been achieved where past data existed.

The performance of the Boris push has been found to be the fastest in terms of pure computational cost. The magnetic gyration of particles is best studied via the Boris push only. However, the cross-field drift performance is found to be generally better for the Vay integrators. Thus, it is recommended that, barring computational limitations, one resort to the latter for studying cross-field drifts. The HC push, though it fares significantly better in terms of phase lag developed over the course of the simulation time, is found to be costly for the accuracy that is achieved. Indeed, the performance is most often comparable to, if not worse than, the other two pushers. The authors recommend sticking to the Boris and Vay pushers as and when required.

Finally, a fitness parameter $f$ has been proposed to quantify the computational cost per unit error developed in each iteration. This has been designed to prioritize both accuracy and computational cost. A higher $f$ does not necessarily mean a better integrator, and neither is the reverse true. Furthermore, the $f$ parameter is a purely comparative number and there is no absolute range of "goodness" or "badness". It all depends on the system being studied. A more comprehensive study of how $f$ varies for different systems can be studied in the future.

\section*{Acknowledgements}
M. Yasir would like to acknowledge the Indian Institute of Technology Delhi for providing the financial grant.

\section*{Conflict of Interest}
The authors of this paper have no conflicts to disclose

\section*{Data Availability}
The data that support the findings of this study are available with the corresponding author upon reasonable request.

\bibliographystyle{unsrt}
\bibliography{bibliography}

\end{document}